\documentclass{PoS}

\title{A Short History of Spin}

\ShortTitle{A Short History of Spin}

\author{\speaker{Richard G. Milner}%
         \thanks{Chair, International Spin Physics Committee, 2013-2017.}\\
        Laboratory for Nuclear Science, MIT, Cambridge, MA 02139\\
        E-mail: \email{\bf milner@mit.edu}}

\abstract{The paper outlines the historical development of spin in physics from about 1920 to the present day.  It aims to provide the student with an accurate chronology of important developments, both scientific and technical.}

\FullConference{XVth International Workshop on Polarized Sources, Targets, and Polarimetry\\
                 September 9 - 13, 2013\\
                 Charlottesville, Virginia, USA}

\begin{document}

\section{Introduction}
Classical mechanics, specifically rigid body motion, contains the ideas of spin and its relationship to angular momentum.  For example, in astronomy spin-orbit coupling between celestial objects reflects the conservation law of angular momentum.  Around 1920, the study of atomic systems led to a failure of classical mechanics and the emergence of an entire new paradigm to describe the subatomic world, namely quantum mechanics.   Spin was found to be an essential quantum number of all subatomic particles and its profound properties underpin the physicist's present-day understanding of the structure of matter. In this paper, I provide a brief outline of the history of spin from 1925 to the present day both in terms of the scientific understanding of the fundamental structure of matter as well as the development of key experimental tools.  I have found it useful to relate the history as a narrative involving four consecutive time periods. 

\section{Explaining the Fundamental Structure of Matter: 1925-50}

In the early 20th century it became evident that atoms and molecules with even numbers of electrons are more chemically stable than those with odd numbers of electrons. For example, Lewis in the third of his six postulates of chemical behavior stated~\cite{Lewis1916} that the atom tends to hold an even number of electrons in the shell which surrounds the nucleus. In 1919, Langmuir suggested~\cite{Langmuir1919} that the periodic table could be explained if the electrons in an atom were connected or clustered in some manner.  

About 1920, the Bohr-orbit theory provided the accepted understanding of the atom by physicists.  However, the number of states observed experimentally was double what was predicted by Bohr-Sommerfeld quantization rules.  This mysterious doubling was known as {\it Mechanische Zweidentigkeit} in German and as {\it duplexity} in English.  In 1922, Stern and Gerlach carried out their famous experiment~\cite{Stern1922} of passing silver atoms through an inhomogeneous magnetic field and observing a deflection either up or down by the same amount.   At the time, the experiment was interpreted as a crucial validation of the Bohr-Sommerfeld theory over the classical theory of the atom.  It showed clearly that spatial quantization exists, a phenomenon that can be accommodated only within a quantum mechanical theory.  In 1924, Bose, at Dhaka University, derived Planck's quantum radiation law by counting states with identical properties and sent it to Einstein, who translated it into German himself, and had it published in Zeitschrift f\" ur Physik~\cite{Bose1924}.  In 1925, the Pauli Exclusion Principle was formulated~\cite{Pauli25} as: {\it no two electrons can have identical quantum numbers}.  

Most significantly for this discussion, also in that year, Leiden graduate students Uhlenbeck and Goudsmit first hypothesized~\cite{Uhl25} intrinsic spin as a property of the electron.    This occurred over the strong objections of some prominent physicists but with the support of their advisor, Paul Ehrenfest.  In their Nature letter they write: {\it "It seems that the introduction of the concept of the spinning electron makes it possible throughout to maintain the principle of the successive building up of atoms utilized by Bohr in his general discussion of the relations between spectra, and the natural system of the elements.  Above all, it may be possible to account for the important results arrived at by Pauli without having to assume an unmechanical duality in the binding of the electrons."}  In the succeeding letter in the same journal, Bohr fully agreed. 

In 1926, Thomas~\cite{Thomas1926} correctly applied relativistic calculations to spin-orbit coupling in atomic systems and resolved a missing factor of two in the derived {\it g}-values.  Also in 1926, Fermi~\cite{Fermi1926} and Dirac~\cite{Dirac1926} developed the Fermi-Dirac statistics for electrons.  It was immediately applied to describe stellar collapse to a white dwarf~\cite{Fowler1926}, to electrons in metals~\cite{Sommerfeld1927}, and to field electron emission from metals~\cite{Fowler1928}.

In 1928, Dirac developed~\cite{Dirac28} his elegant equation for spin-$\frac{1}{2}$ particles.  In this formulation, solutions are four-component spinors which are interpreted as positive and negative energy states of spin $\pm \frac{1}{2}$ each.   Dirac predicted the existence of the positron, and the theory became the basis for the most precisely tested theory in physics, Quantum Electrodynamics.  By the end of the 1920s, physicists had developed a fundamental understanding of the essential role of electron spin in explaining the electronic structure of the atom.  There exist excellent, personal, historical accounts by Dirac~\cite{Dirac1974}, Uhlenbeck~\cite{Uhlen1976}, and Goudsmit~\cite{Gouds1976} of this period.

In 1927, Wrede, a student of Stern at Hamburg~\cite{Wrede1927}, and Phipps and Taylor at Illinois~\cite{Phipps1927} independently observed the  deflection of atomic hydrogen in a magnetic field gradient.  In  1929, Mott wondered~\cite{Mott1929} if electron spin can be observed directly via the scattering of electrons from atomic nuclei.  Note that in the Appendix to his paper, Mott showed that the Stern-Gerlach experiment cannot be carried out for electrons.  Only in 1942 did Shull {\it et al.} verify~\cite{Shull1943} Mott's prediction in a double scattering experiment which used 400 keV electrons from a Van de Graaf generator.  In the mid-1920s, Heisenberg and Hund postulated the existence of two kinds of molecular hydrogen: {\it orthohydrogen} where the two proton spins are aligned parallel and {\it parahydrogen} where the two proton spins are antiparallel.  By the end of the decade, they had been studied experimentally.  Later, by deflection of orthohydrogen in a magnetic field gradient, Stern and collaborators measured the {\it g}-factor of the proton to be about 2.5 nuclear magnetons~\cite{Estermann1933}, a marked deviation from the Dirac value for a pointlike spin-$\frac{1}{2}$ particle, and the first hint of its internal structure.

In the 1930s, Rabi and collaborators (inc. N. Ramsey and J. Zacharias) using molecular beams in a weak magnetic field measured the magnetic moments and nuclear spins of hydrogen, deuterium, and heavier nuclei~\cite{Kellogg1939}.

By the end of the 1940s, the nuclear shell model had been established~\cite{Mayer1948}.  This explained the properties and structure of atomic nuclei and underscored the essential role of proton and neutron spin.  A key aspect was the strong role of spin-orbit coupling, which was suggested to Goeppert-Mayer by a question from Fermi.

\section{Developing Spin as an Experimental Tool: 1950-75}

By the middle of the twentieth century, the intrinsic spin of subatomic particles was a cornerstone of the physicist's theoretical understanding of the fundamental structure of matter.  However, spin as an experimental tool became a 
reality only in this second era.  In the 1950s, a number of seminal experiments were carried out using spin. In 1956, Lee and Yang pointed out that parity should be violated in the weak interaction~\cite{YangLee1956}. 
Shortly afterwards, in 1956, Wu and collaborators observed~\cite{Wu1957} parity violation in aligned $^{56}$Co. In 1958, it was shown experimentally~\cite{Goldhaber1958} using polarization techniques that the neutrino has negative helicity.  1959, the Thomas-Bargmann-Michel-Telegdi equation describing the spin precession of an electron in an external electromagnetic field was derived~\cite{BMT1959}.

In the 1960s, the discovery of pointlike constituents in the proton at SLAC using deep inelastic scattering (DIS) profoundly affected our understanding of the fundamental structure of matter.  A key determination that these constituents had spin-$\frac{1}{2}$ led to their identification as the quarks of SU(3) symmetry.  Important sum rules related to spin-dependent DIS were derived by Bjorken~\cite{Bjorken1966} and by Ellis and Jaffe~\cite{Ellis1974}.  

During this period, the international spin community grew significantly in size to become the active, subfield of international physics we have to-day.  Beginning in 1960 at Basel, symposia on polarization phenomena in nuclear reactions were held every 5 years until 1994.  Beginning in 1974 at Argonne, symposia on high energy spin physics were held every 2 years until 1994.  Beginning in 1996 in Amsterdam, the international spin community became unified and a symposium on spin physics has been held every two years since then.  The International Spin Physics Committee was formed to oversee the organization of this biennial symposium which took place most recently in Dubna, Russia in 2012.  The next meeting is scheduled to take place in Beijing, China in 2014.
The published proceedings of these meetings form the essential record of the research activities over this time.  In~\cite{Schieck}, there is a complete tabulation of these meetings as well as references to their proceedings.  Further, important conventions at Basel in 1960 for spin-$\frac{1}{2}$ particles~\cite{Basel} and at Madison in 1970 for spin-$1$ particles~\cite{Madison} were established to facilitate consistent discussion of spin observables. 
In this period of twenty five years, enormous progress was made in developing spin as an experimental tool.   Highlights are summarized under a number of important developments as follows:

\subsection{Polarized ion sources}
This brief summary is based on the historical reviews by Haeberli in 1967~\cite{Haeberli1967} and in 2007~\cite{Haeberli2008}.  In 1946, Schwinger suggested~\cite{Schwinger1946} using double scattering of neutrons to determine the sign of the spin-orbit coupling. Wolfenstein pointed out~\cite{Wolfenstein1946} that using protons was more practical.  The first experiment was carried out by Heusinkveld and Freier~\cite{Heus1952} to resolve the Fermi-Yang ambiguity~\cite{Bethe1955}, which arises in the analysis of the scattering between particles of 
spin-$\frac{1}{2}$ and spin 0. 

The atomic beam source (ABS) to produce polarized atoms became a reality in this period.  In 1951, Paul proposed to use magnetic multipoles to focus atomic beams.  Experimental work directed at the preparation of polarized-ion beams for nuclear experiments by the atomic-beam method was first undertaken by Clausnitzer, Fleischmann, and Schopper in 1956~\cite{Clausnitzer1956}.  Radiofrequency transitions were first developed at Saclay where the atomic beam method was to involve ionization in the cyclotron magnetic field.  Generally, they utilized the  {\it adiabatic fast passage method}, as proposed by Abragam and Winter~\cite{Abragam1958}.  By use of RF transitions, one can freely choose between vector and tensor deuteron polarizations.  The development of the atomic beam source got underway at Erlangen in 1958 by Fleischmann's group.  Further, in 1964, Gruebler, Schwandt, and Haeberli developed~\cite{Gruebler1964} the first source of polarized H$^-$.  The ABS produced an intensity of $\approx 10^{16}$ /sec with high polarization ($\approx$ 90\%)  by 1970.   In addition, polarized proton sources using the hydrogen metasable state were invented in the 1960s~\cite{Haeberli1967}.

Further, deuteron sources played an important role in spin physics.  The first operational polarized ion source was the Basel deuteron source which operated in 1960 already. Measurement of the first asymmetries was carried out using a polarized beam from an ABS. In contrast to the Erlangen approach, Huber's group at Basel reduced the background signal by using deuterons. 
In addition, in 1975, there was already a polarized $^6$Li beam at the Heidelberg tandem accelerator which was used for scattering experiments. Later, $^7$Li and $^{23}$Na beams were used at tandems to measure reactions including sub-barrier fusion of aligned $^{23}$Na ions. 

\subsection{Polarized electron sources}

The brief summary here is based on the historical review of the development of polarized electrons sources by Prescott~\cite{Prescott2006}.
In 1963, Hughes and colleagues begin consideration of polarized electron sources at Yale~\cite{Long1965}.  The initial work was based on production of polarized electrons
by photoionization of a polarized atomic beam of alkali atoms.  This effort produced the PEGGY source based on photoionization of $^6$Li  which was commissioned at SLAC in 1974.
In 1971, Sokolov-Ternov self-polarization~\cite{Sokolov1963} was first observed with the 625 MeV electron beam at the VEPP-2 storage ring, Novosibirsk, Russia.
Further, a source of polarized electrons was developed~\cite{heliumafterglow} using optically oriented metastable atoms in a flowing helium afterglow.

\subsection{Polarized proton targets}
The brief summary here is based on the 2013 review of Keith~\cite{Keith2013}.  While the earliest examples of polarized targets were polarized by static methods (i.e. low temperatures and high magnetic fields, see for example~\cite{Bernstein1954}), dynamic nuclear polarization (DNP) has been used for the majority of solid polarized targets used in nuclear and high energy physics. This technique began in 1953, when Overhauser at Illinois proposed to transfer the polarization of conduction electrons in a metal to nuclei by saturating the electron's spin resonance with RF radiation~\cite{Overhauser1953}.  Initially met with great skepticism, Overhauser's suggestion was experimentally verified by Carver and Slichter later that year~\cite{Carver1953}.  Working independently, Jefferies~\cite{Jefferies1957} and Abragam~\cite{Abragam1962} both suggested to dynamically polarize nuclei by saturating so-called "forbidden transitions" in which electron and nuclear spins flip simultaneously.  In 1962, Abragam, Borghini and co-workers built the first polarized proton target for the 20 MeV polarized proton beam at Saclay~\cite{Abragam1962}.  Shortly thereafter, Chamberlain, Jefferies, and collaborators built a polarized proton target for 250 MeV pion scattering experiments at Berkeley~\cite{Chamberlain1963}.  In both cases the average polarization of protons in the LMN target material was only about 20\%, but steady improvement would be made in the following years, thanks to refinements in magnets and cryogenics~\cite{Borghini1970}.

In the late 1960's, efforts were made to develop target materials with a higher percentage of polarized free protons (compared to the 3\% in LMN).  This work was led in part by Borghini at CERN and motivated by advances in the spin-temperature description of the DNP process~\cite{Borghini1971}.  It culminated in 1974, when de Boer and Niinikoski demonstrated 98\% proton polarization in propanediol doped with the Cr(V) free radical~\cite{deBoer1974}.  By the early 1970's, diols and alcohols had replaced LMN as the target material of choice, due to their higher free proton content, higher resistance to radiation damage, and higher polarizations.   Around the same time, the first highly successful frozen spin target was built at CERN by Niinikoski and Udo~\cite{Niinikoski1976} and utilized advances that Niinikoski had made in $^3$He-$^4$He dilution refrigeration~\cite{Niinikoski1971}.  In 1979, Niinikoski and Rieubland produced proton polarizations greater than 90\% in irradiated NH$_3$~\cite{Niinikoski1979}, which would soon become a common target material (especially for higher luminosity experiments), thanks to its even higher proton content and radiation resistance~\cite{Meyer2004}.  Further details about the current status of polarized proton targets can be found in the comprehensive review by Crabb and Meyer~\cite{Crabb1997}.

Note that later development work, see section 4.3 below, produced highly polarized hydrogen, deuterium, and $^3$He gas targets for
use with stored beams in storage rings.
Finally, it must be noted that in the early 1970s, proton spin was proposed as a diagnostic tool for medicine. This evolved into the technique of Magnetic Resonance Imaging, now in use daily worldwide.

\subsection{Polarized $^3$He beams and targets}

Important spin capabilities using spin-1 photons were invented in this period.  In 1950, Kastler proposed~\cite{Kastler1950} the technique of optical pumping and, in 1960, the laser was developed.    The technique to polarize $^3$He gas using metastability exchange optical pumping was developed by Colegrove, Schearer, and Walters in 1963~\cite{Colegrove1963}.   Polarized $^3$He sources and targets based on flash lamps were used in nuclear physics experiments.  For example, the spin-dependence of the fusion cross section $^2$H+$^3$He $\rightarrow$ $^4$He+$^1$H was measured at Basel in 1971~\cite{Leemann1971}.   Magnetometers using polarized $^3$He were developed and located beneath the oceans to detect submarines.

\section{Using the Tool: 1975-95}

In this period, the tools invented in the previous era started to be employed to great effect.  In particular, experiments using polarized beams at high energies at ANL, CERN, Fermilab and SLAC produced results which continue to shape our understanding of the fundamental structure of matter.   At SLAC, the development of polarized electron sources initiated a program of measurements which continued until the end of the century.  At CERN, the highly polarized, high energy muon beams resulting from inflight pion decay have resulted in a series of experiments EMC, SMC, and COMPASS which have used DNP targets and have profoundly impacted our understanding of the structure of the nucleon.  In addition at lower energies, polarized beams played a central rule in experimental studies, {\it e.g.} spectroscopy (resolving ambiguities in angular momentum assignments), determination of spin-dependence in the optical model,  constraining the effects of the D-state in light nuclei~\cite{Weller1988}, and angular momentum dependence in stripping and pickup reactions.

\subsection{High energy lepton scattering}
The 1970s saw the SLAC experiments E80 and E130 measure spin-dependent inclusive DIS from the proton for the first time.  It was found that the valence quarks in the proton were polarized as expected~\cite{Hughes1983}.  At SLAC plans to construct the SLC were underway, so a subsequent proposal led by Hughes to probe the valence quark region in the neutron was turned down.  Hughes turned his attention to CERN where the EMC experiment was being mounted.  It took data for about a decade and in 1988 produced spin-dependent DIS data on the proton at low $x$ for the first time.  The EMC data showed~\cite{Ashman1988} that the quarks carried far less of the proton's spin than was expected and that the Ellis-Jaffe Sum Rule was violated. Thus, the "proton spin crisis" was born. 
  
Also at SLAC, experiment E122 for the first time used a polarized electron source based on optical pumping of GaAs.  This technology dramatically enhanced the ability to carry out experiments using polarized electron beams.  At that time, the electron beam polarization was limited to about 40\%.  In 1978, E122 announced the observation of parity violating electron scattering at SLAC for the first time~\cite{Prescott1978}. This validated the Weinberg-Salam Model and provided the first measurement of the neutral current coupling of the electron.

Following the discovery of our incomplete understanding of the proton's spin, four major campaigns at CERN/SMC/COMPASS, SLAC/End Station A, DESY/HERMES, and the future RHIC-spin were conceived in this period.   In particular, at SLAC, the use of strained GaAs resulted in significantly higher polarization~\cite{Maruyama1991}.  Subsequently, this technology has been employed at MIT-Bates, Jefferson Lab, Mainz, and Bonn.

Finally, at SLAC an important series of measurements with polarized electrons at the Z-pole got underway.  The use of polarization at SLAC allowed the SLC to compete with the significantly higher luminosity LEP experiments in carrying out precision tests of the Standard Model.  At LEP, transverse polarizations via Sokolov-Ternov self-polarization in excess of 50\% were observed and the effects of the Earth's tides on the beam tune were observed~\cite{Assmann}.

\subsection{High energy proton scattering}
In this period, high energy proton-proton spin experiments were carried out at the ZGS, AGS, and Fermilab.  This brief summary is taken from the review article by Krisch~\cite{Krisch2007}.   High energy polarized proton beams became available in 1974 at the Argonne ZGS~\cite{Fernow1981}.  Krisch and collaborators measured unexpectedly large asymmetries at high p$^2_{\perp}$ in pp elastic scattering in contradiction to the expectations from QCD~\cite{OFallon1977}. To this day, these results do not have an accepted explanation.  With the phaseout of the ZGS in the late 1970s, polarized protons were developed at the AGS at Brookhaven National Laboratory.  It was far more difficult to accelerate protons in the strong-focusing AGS than in the weak-focusing ZGS.  To accelerate polarized protons to 22 GeV at the AGS, 45 strong depolarizing resonances were successfully overcome. 

At Fermilab beginning in the 1970s, polarization experiments using protons were carried out.  Inclusive hyperon polarization experiments showed~\cite{Heller1983} increasing polarization with $p_T$.  These data are consistent with other experiments carried out at KEK and the CERN ISR.  Further experiments at Fermilab carried out by Yokasawa and colleagues measured~\cite{Adams1991} $A_N$, the transverse single spin asymmetry, at 200 GeV for inclusive forward $\pi^\pm$ meson production and measured increasing asymmetries vs. $x_F$.   $A_N$ measurements vs. $x_F$ at $\sqrt{s}$ = 4.9 GeV at ANL~\cite{Klem1976}, at $\sqrt{s}$ = 6.6 GeV at BNL~\cite{Allgower2002},  at $\sqrt{s}$ = 19.4 GeV at Fermilab~\cite{Adams1991}, and at $\sqrt{s}$ = 62.4 GeV at RHIC~\cite{Arsene2008} all  show a striking similarity.

In the 1980s, a new 20 TeV on 20 TeV proton-proton collider was being planned in the U.S.  Each 20 TeV SSC ring would have about 36,000
resonances.   It was concluded that it should be possible to accelerate and maintain the polarization of 20 TeV protons in the SSC, only if the new Siberian-snake (named by Courant) concept of Derbenev and Kondratenko~\cite{Derbenev1978} really worked.  This motivated the development at IUCF.   Twenty six empty spaces for Siberian snakes were added in each SSC ring.  The SSC was cancelled in 1993. 

\subsection{Spin in storage rings}
During this period, considerable progress was made in the manipulation of spin in storage rings.   In 1989, the Siberian Snake concept was demonstrated for the first time at IUCF~\cite{Krisch1989} and plans proceeded to implement this technology in RHIC at BNL.  In 1984, construction of the HERA electron-proton collider got underway with planned polarized e$^-$/e$^+$ beams using Sokolov-Ternov polarization and Richter-Schwitters spin-rotators~\cite{Richter1974}.  In 1988, the HERMES experiment was proposed and, subsequently in 1994, significant lepton polarization was observed in HERA at 27 GeV via the Sokolov-Ternov mechanism for the first time.

The development of intense ($\approx 10^{17}$ /sec), highly polarized ($\approx 95\%$) atomic beams to feed storage cells made substantial progress in the 1980s~\cite{Steffens2003}. In addition, a new generation of powerful lasers enabled the development of optically pumped sources of polarized hydrogen, deuterium and $^3$He at ANL, Caltech, Harvard, and Princeton~\cite{Chupp1994}.  Work at Madison and Heidelberg on ABS fed internal gas targets was focused on a measurement of spin-filtering at the Heidelberg TSR~\cite{Stock1994}.  At Novosibirsk, pioneering internal target experiments were carried out in collaboration with the ANL group~\cite{Novosibirsk}. 

\subsection{Spin at medium energy accelerators}
In this period, a number of medium energy accelerators using polarized beams came online.  Many were located at research universities and attracted new generations of young experimentalists into spin physics. These included Saclay in France, NIKHEF in the Netherlands, Mainz and Bonn in Germany, TRIUMF in Canada, Novosibirsk in Russia, and MIT-Bates and IUCF in the U.S.  With the decision to construct a new, CW, multi-GeV electron accelerator in the U.S., significant theoretical work began on electromagnetic nuclear physics.  In particular, Donnelly and colleagues developed~\cite{Donnelly1986} the {\it SuperRosenbluth} technique, where spin was used as an experimental  "knob" to maximize sensitivity to particularly important pieces of physics.  This thinking had a major impact on the design of medium energy electron scattering experiments for the next several decades. 

Scientific highlights from this period include: the first measurement in 1984 at MIT-Bates of t$_{20}$ in elastic electron-deuteron scattering to allow separation of the three elastic form factors of the deuteron~\cite{The1991};  tests of the Standard Model via measurement of parity-violating quasielastic electron scattering from beryllium at Mainz~\cite{Heil1989} and elastic electron scattering from $^{12}$C at Bates~\cite{Souder1990}; and the first measurements of spin-dependent electron scattering from polarized $^3$He gas targets at MIT-Bates~\cite{BatesHe3}.   At IUCF in 1993, the first experiment to use both polarized beam (proton) and polarized internal gas target ($^3$He) was carried out~\cite{Lee1993} to study the spin structure of $^3$He .  The MIT-Bates and IUCF measurements showed that polarized $^3$He could be used as an effective polarized neutron target for scattering experiments and set the stage for subsequent lepton scattering experiments at SLAC, DESY, Mainz, and Jefferson Lab.

Polarimetry, both for polarized beams, and for recoil particles, made enormous strides in this period.  For electron beams, both Mott and M\" oller scattering were used as well as laser backscattering.  Recoil neutron and proton polarimeters were developed to measure the nucleon elastic form-factors in the transfer of polarization from incident polarized electron beam on unpolarized targets. Subsequently, they have been used to great effect in the Jefferson Lab program.

\subsection{Hadronic parity violation}
Measurement of hadronic parity violation using spin observables became an active field of study in this period: see reviews in 1985~\cite{Adelberger1985}, in 1988~\cite{ProcHadPV1988} and in 2013~\cite{Haxton2013}.  The theoretical underpinning for this experimental program was the 1980 meson-exchange approach of Desplanques, Donoghue, and Holstein~\cite{DDH1980}. The analyzing power for the scattering of longitudinally polarized protons from an unpolarized proton target was measured at 13.6 MeV~\cite{Eversheim1991} and 15 MeV~\cite{Nagle1978} by groups from Bonn and Los Alamos, and at 45 MeV~\cite{Balzer1980} by a group from PSI. Further, a medium energy measurement at 220 MeV was made at TRIUMF~\cite{Berdoz2001}.  The TRIUMF experiment importantly used an optically pumped polarized proton source~\cite{Levy1996}.  The measured asymmetry is typically of order 10$^{-7}$.  Further experiments to find parity violation in $np \rightarrow d \gamma$ at Grenoble and LANL and in the circular polarization of 2.22 MeV photons emitted in the capture of unpolarized thermal neutrons by protons at St. Petersburg were inconclusive.  As so few of the experiments on NN and few body systems succeeded in isolating nonzero effects, experimenters turned to nuclei where parity violating observables might be enhanced.  Several such experiments yielded nonzero results.

\section{Spin in Use Worldwide: 1995-present}

Beginning in the mid-1990s, major experiments in spin physics at accelerators worldwide began to come online.  They were built on the technical developments of previous decades and were motivated by important scientific questions.  These included: the origin of nucleon spin; the use of the SuperRosenbluth technique to measure the elastic proton and neutron form-factors; and tests of the Standard Model.

For a modern review of the spin structure of the nucleon, see Aidala {\it et al.}~\cite{Aidala2013}.  The HERMES experiment took data from 1995 through 2007, when operation of HERA ceased.   It used the self-polarized 27 GeV electron and positron beams of the HERA ring incident on polarized internal gas targets of hydrogen, deuterium~\cite{Airapetian2005}, and $^3$He~\cite{DeSchepper1998}.  For the first time, it employed sophisticated hadron detection and measured semi-inclusive DIS.  HERMES provided first measurements of a number of important effects accessible in SIDIS or exclusive measurements.  This has been subsequently pursued at Jefferson Lab and CERN/COMPASS and has underpinned the modern theoretical framework of nucleon structure based on multivariable Wigner distributions. 

The COMPASS experiment continued the successful CERN polarized 160 GeV muon fixed target program into the new century.    Both targets and detectors were substantially upgraded to include both longitudinal and transverse proton and deuteron as well as a RICH detector for particle identification.   COMPASS provides unique data~\cite{Martin2011} at higher momentum transfer complementary to data from HERMES and Jefferson Lab. 

With electron beam polarizations of over 80\%, a new generation of experiments to measure inclusive DIS on polarized proton, deuteron, and $^3$He targets in End Station A at SLAC yielded precision data on the spin-dependent structure functions~\cite{Anthony2003, Anthony1996}. Targets based on atmosphere volumes of polarized $^3$He were realized for the first time using spin exchange optical pumping~\cite{Anthony1996} and used with 10 $\mu$A of electron beam.  This technology has subsequently been further developed and employed to great effect at Jefferson Lab.  

In the early part of this period, successful spin physics programs were carried out at the medium energy accelerators NIKHEF, IUCF, and MIT-Bates.  At NIKHEF and Bates, stored electron beams of energy 600-850 MeV reached intensities of 250 mA and polarizations of about 65\%.  They were used with internal polarized gas targets to measure spin-dependent electron scattering from polarized hydrogen and deuterium~\cite{Hasell2011}.  A scientific highlight from this work was the determination of the neutron elastic form-factor at low momentum transfer with precision comparable to that of the proton~\cite{Geis2008}.   At IUCF, the PINTEX program made precision measurements of spin-dependent proton-proton and proton-deuteron scattering between 100 and 500 MeV~\cite{Przewoski2006}.  Also at IUCF, evidence for a three-nucleon force in spin-dependent elastic proton-deuteron scattering was obtained using a laser driven polarized deuterium target~\cite{Cadman2001}.  At both Bates and IUCF, highly efficient spin reversal of the stored beam was effected using an RF solenoid magnet~\cite{Blinov1998, Morozov2001} .  Regrettably, by 2005 the accelerators at all three laboratories had ceased operation.  Fortunately, the COSY accelerator, with its polarized stored beams and polarized targets, remains an active laboratory for spin physics at medium energies~\cite{Rathmann2010}.

Although polarized beams were not in the original scope, by 1998 Jefferson Lab was operating at 4~GeV with intense, polarized electron beams.  Spin measurements are central to Jefferson Lab's scientific program and the spin technical capabilities in terms of beams, targets, and polarimeters are world class.  One of the major scientific results from Jefferson Lab was the discovery that the proton elastic form-factor ratio was dramatically different when measured with the recoil polarization technique compared to that obtained using the unpolarized cross section~\cite{Puckett2012}.  The widely accepted explanation is that this is due to contributions beyond single photon exchange in the QED expansion.   

Analysis of spin-dependent DIS data raised the possibility that the {\it strange} quarks carried unexpectedly large polarization.  This motivated a worldwide program of parity violating electron scattering at MIT-Bates, Mainz, and Jefferson Lab to look for strange quark contributions to the proton's magnetic moment and charge~\cite{Beck2001}.  While these experiments drove the technical development of polarized electron beams at these labs, by 2010 it was concluded that there were no large contributions from {\it strange} quarks~\cite{Armstrong2012}.

The RHIC accelerator turned on in 2000 with heavy ion beams and by 2006 the world's first polarized proton collider had been realized there.  
RHIC-spin employs a high intensity (3-5 mA),  optically pumped $>$80\% polarized proton source as well as Siberian Snakes and requires careful spin manipulation through multiple accelerators to achieve 60\% polarized colliding proton beams at center-of-mass energies up to 500 GeV.  
It was made possible with substantial collaboration and investment from Japan.   By 2011, the first measurement of parity-violating W-boson production was carried out~\cite{Aggarwal2011}.  In 2013, it was concluded that the contributions of gluons to the proton spin was comparable to that of the quarks~\cite{Surrow2013}.  

Polarization continues to be an essential experimental technique to test fundamental symmetries and to search for new physics beyond the Standard Model. At SLAC, experiment E158 carried out a precision test of the Standard Model by measuring the Weinberg angle in spin-dependent M\o ller scattering~\cite{Anthony2005}.  At BNL, E821 has reported a value for the muon's anomalous magnetic moment which challenges the Standard Model~\cite{Bennett2006}.
The availability of low-energy, polarized neutron beams for fundamental research has been realized.  Polarized ${}^3$He neutron spin filters are widely used~\cite{Huber2009} in this work.  Searches for non-vanishing electric dipole moments (EDMs) of fundamental particles  implies violation of both parity and time reversal symmetries.  Several experiments to look for non-zero neutron EDM are under development~\cite{Ito2007}.  Searches for non-zero EDMs of the proton and light nuclei in storage rings are also actively being considered~\cite{Lenisa2013}.  Many exciting, open questions remain in spin physics~\cite{Jaffe2003}.
\newpage
\section{Acknowledgements}
I would like to thank Don Crabb and the organizers of PSTP2013 for motivating me to write this paper.  I  thank Willy Haeberli, Chris Keith, Alan Krisch, Matt Poelker, Charles Prescott, and Erhard Steffens for providing me with essential information and insight into the history of spin.  Due to space and time constraints, my effort to describe the ninety year history of such a large field of scientific endeavor is necessarily incomplete and is influenced by my personal expertise and interests.  Suggestions to ameliorate these limitations are welcome. My research is supported by the Office of Nuclear Physics at the United States Department of Energy.

\end{document}